\documentclass[twocolumn,tighten]{aastex63}
\usepackage{privon_astro}
\usepackage{xcolor}
\newcommand\crule[3][black]{\textcolor[rgb]{#1}{\rule{#2}{#3}}}

\newcommand{\myemail}{george.privon@ufl.edu}

\newcommand{\LHX}{L$_{3-24~\rm keV}$\xspace}
\newcommand{\LAGN}{L$_{\rm AGN}$\xspace}
\newcommand{\LBOL}{L$_{\rm bol}$\xspace}
\newcommand{\NH}{N$_{\rm  H}$\xspace}

\newcommand{\nustar}{{\em NuSTAR}\xspace}
\newcommand{\chandra}{{\em Chandra}\xspace}

\shortauthors{Privon et al.}

\begin{document}

\title{A Hard X-ray Test of HCN Enhancements As a Tracer of Embedded Black Hole Growth}
\email{\myemail}

\author[0000-0003-3474-1125]{G. C. Privon}
\affiliation{Department of Astronomy, University of Florida, P.O. Box 112055, Gainesville, FL 32611, USA}

\author[0000-0001-5231-2645]{C. Ricci}
\affiliation{N\'ucleo de Astronom\'ia de la Facultad de Ingenier\'ia, Universidad Diego Portales, Av. Ej\'ercito Libertador 441, Santiago, Chile}
\affiliation{Kavli Institute for Astronomy and Astrophysics, Peking University, Beijing 100871, China}
\affiliation{George Mason University, Department of Physics \& Astronomy, MS 3F3, 4400 University Drive, Fairfax, VA 22030, USA}

\author{S. Aalto}
\affiliation{Department of Space, Earth and Environment, Chalmers University of Technology, Onsala Space Observatory, SE-439 92, Onsala, Sweden}

\author[0000-0001-8504-8844]{S. Viti}
\affiliation{Department of Physics and Astronomy, University College London, Gower Street, London, WC1E 6BT, UK}

\author{L. Armus}
\affiliation{IPAC, MC 314-6, Caltech, 1200 E. California Blvd., Pasadena, CA 91125, USA}

\author{T. D\'iaz-Santos}
\affiliation{N\'ucleo de Astronom\'ia de la Facultad de Ingenier\'ia, Universidad Diego Portales, Av. Ej\'ercito Libertador 441, Santiago, Chile}
\affiliation{Chinese Academy of Sciences South America Center for Astronomy, National Astronomical Observatories, CAS, Beijing 100101, China}

\author{E. Gonz\'{a}lez-Alfonso}
\affiliation{Universidad de Alcal\'{a}, Departamento de F\'isica y Matem\'aticas, Campus Universitario, E-28871 Alcal\'a de Henares, Madrid, Spain}

\author[0000-0002-4923-3281]{K. Iwasawa}
\affiliation{Institut de Ci\`encies del Cosmos (ICCUB), Universitat de Barcelona (IEEC-UB), Mart\'i i Franqu\`es, 1, E-08028 Barcelona, Spain}
\affiliation{ICREA, Pg. Llu\'is Companys 23, E-08010 Barcelona, Spain}

\author[0000-0003-0416-4830]{D. L. Jeff}
\affiliation{Department of Astronomy, University of Florida, P.O. Box 112055, Gainesville, FL 32611, USA}

\author{E. Treister}
\affiliation{Instituto de Astrof\'sica, Facultad de F\'isica, Pontificia Universidad Cat\'olica de Chile, Casilla 306, Santiago 22, Chile}

\author{F. Bauer}
\affiliation{Instituto de Astrof\'sica, Facultad de F\'isica, Pontificia Universidad Cat\'olica de Chile, Casilla 306, Santiago 22, Chile}
\affiliation{Millennium Institute of Astrophysics (MAS), Nuncio Monse{\~ n}or S{\'o}tero Sanz 100, Providencia, Santiago, Chile}
\affiliation{Space Science Institute, 4750 Walnut Street, Suite 205, Boulder, Colorado, 80301, USA}

\author{A. S. Evans}
\affiliation{Department of Astronomy, University of Virginia, P.O. Box 400325, Charlottesville, VA 22904, USA}
\affiliation{National Radio Astronomy Observatory, 520 Edgemont Road, Charlottesville, VA 22903, USA}

\author[0000-0002-5923-2151]{P. Garg}
\affiliation{Department of Astronomy, University of Florida, P.O. Box 112055, Gainesville, FL 32611, USA}

\author[0000-0002-7758-8717]{R. Herrero-Illana}
\affiliation{European Southern Observatory (ESO), Alonso de C\'ordova 3107, Vitacura, Casilla 19001, Santiago de Chile, Chile}
\affiliation{Institute of Space Sciences (ICE, CSIC), Campus UAB, Carrer de Magrans, E-08193 Barcelona, Spain}

\author[0000-0002-8204-8619]{J. M. Mazzarella}
\affiliation{IPAC, MC 100-22, California Institute of Technology, Pasadena, CA 91125, USA}

\author{K. Larson}
\affiliation{Department of Astronomy, California Institute of Technology, 1200 E. California Blvd., Pasadena, CA 91125, USA}

\author{L. Blecha}
\affiliation{Department of Physics, University of Florida, P.O. Box 118440, Gainesville, FL 32611, USA}

\author{L. Barcos-Mu\~noz}
\affiliation{National Radio Astronomy Observatory, 520 Edgemont Road, Charlottesville, VA 22903, USA}

\author[0000-0002-2688-1956]{V. Charmandaris}
\affiliation{Department of Physics, University of Crete, GR-71003, Heraklion, Greece}
\affiliation{Institute of Astrophysics, Foundation for Research and Technology-Hellas, GR-71110, Heraklion, Greece}

\author{S. Stierwalt}
\affiliation{Infrared Processing and Analysis Center, MC 314-6, Caltech, 1200 E. California Blvd., Pasadena, CA 91125, USA}
\affiliation{Department of Physics, Occidental College, Los Angeles, CA 90041, USA}

\author{M. A. P{\'e}rez-Torres}
\affiliation{Instituto de Astrofísica de Andalucía - CSIC, P.O. Box 3004, E-18008, Granada, Spain}

\begin{abstract}
Enhanced emission from the dense gas tracer HCN (relative to HCO$^+$) has been proposed as a signature of active galactic nuclei (AGN).
In a previous single-dish millimeter line survey we identified galaxies with HCN/HCO$^+$ (1--0) intensity ratios consistent with those of many AGN but whose mid-infrared spectral diagnostics are consistent with little to no ($\lesssim15\%$) contribution of an AGN to the bolometric luminosity.
To search for putative heavily obscured AGN, we present and analyze \nustar hard X-ray (3--79 keV) observations of four such galaxies from the Great Observatories All-sky LIRG Survey.
We find no X-ray evidence for AGN in three of the systems and place strong upper limits on the energetic contribution of any heavily obscured (N$_{\rm H}>10^{24}$ \cmcol) AGN to their bolometric luminosity.
The upper limits on the X-ray flux are presently an order of magnitude below what XDR-driven chemistry models predict are necessary to drive HCN enhancements.
In a fourth system we find a hard X-ray excess consistent with the presence of an AGN, but contributing only $\sim3\%$ of the bolometric luminosity.
It is also unclear if the AGN is spatially associated with the HCN enhancement.
We further explore the relationship between HCN/HCO$^+$ (for several $\mathrm{J}_\mathrm{upper}$ levels) and \LAGN/\LIR for a larger sample of systems in the literature.
We find no evidence for correlations between the line ratios and the AGN fraction derived from X-rays, indicating that HCN/HCO$^+$ intensity ratios are not driven by the energetic dominance of AGN, nor are they reliable indicators of ongoing SMBH accretion.
\end{abstract}


\section{Dense Molecular Gas Tracers and Active Galactic Nuclei}
\label{sec:introduction}

The molecular gas tracers HCN and HCO$^+$ have been implicated as a probe of the dense molecular gas associated with star formation \citep{Solomon1992,Gao2004}.
However, targeted studies have found that the host galaxies of some active galactic nuclei (AGN) have enhanced HCN/HCO$^+$ intensity ratios\footnote{In this paper millimeter "line ratio" refers to the intensity ratio.}, compared to starburst galaxies \citep[e.g.,][]{Kohno2003,GraciaCarpio2006,Imanishi2006,Imanishi2007,Krips2008,Imanishi2009,Davies2012,Izumi2016}.
These findings, made in both galaxy-integrated and resolved (resolution 10's--1000's of pc) measurements, led to speculation that HCN enhancements could be a signpost of AGN activity.

More representative studies have found starburst galaxies that have HCN/HCO$^+$ enhancements similar to the enhancements seen in AGN \citep[e.g.,][]{Costagliola2011,Privon2015,SMartin2015,Konig2018,Harada2018}, where the relative starburst--AGN strength has been primarily assessed using the mid-infrared (MIR) $6.2~\mu m$ polycyclic aromatic hydrocarbon (PAH) equivalent width (EQW).
This suggests two possibilities: 1) the HCN/HCO$^+$ enhancements trace an alternate physical process than AGN activity, or 2) these MIR classified starbursts host AGN that are behind obscuring screens sufficiently thick to render MIR diagnostics ineffective.
The former is supported by theoretical work showing that differences in the HCN/HCO+ line intensity ratios can arise due to many different densities, temperatures, radiation fields, and evolutionary states of activity \citep[e.g.,][]{Viti2017}.
The latter scenario is plausible as the necessary optical depths are seen in the ``compact obscured nuclei'' \citep[CONs;][]{Sakamoto2010,Aalto2015a,Falstad2015,Scoville2015} which can have obscuring columns in excess of N$_{\rm H}>10^{25}$ \cmcol and show emission from vibrationally excited HCN molecules (HCN-VIB).

Though the emission from rotational transitions in molecules (and the associated line ratios) are set by excitation, abundance, and/or opacity effects, here we focus on testing the empirical use of HCN enhancements as a tracer of embedded black hole growth.
We have obtained {\it Nuclear Spectroscopic Telescope Array} ({\it NuSTAR}, \citealp{Harrison2013}) hard X-ray (3-79 keV) observations to search for heavily obscured AGN \citep[\NH $\gtrsim 10^{24}$ \cmcol; e.g.,][]{Teng2015,Ricci2016,Ricci2017}.
We begin by discussing the sample and observations (Section~\ref{sec:observations}) and then present the derived properties of the targeted four systems (Section~\ref{sec:xrays}).
We then use a suite of archival millimeter data and \nustar observations to examine the link between dense gas tracers and known AGN energetics (Section~\ref{sec:results}).

\section{Hard X-Ray Survey}
\label{sec:observations}

\subsection{The Sample}
\label{sec:sample}

Here we study four luminous infrared galaxies (LIRGs; \LIR$>10^{11}$ \Lsun) from the Great Observatories All-sky LIRG Survey \citep{Armus2009} and the IRAM 30m dense molecular gas survey of \citet{Privon2015}.
These objects have elevated \HCNj{1}{0}/\HCOj{1}{0} line ratios ($>1.5$) and are classified in the MIR as starburst or composite objects based on the equivalent width (EQW) of their $6.2~\mu m$ polycyclic aromatic hydrocarbon (PAH) emission features \citep{Stierwalt2013}.
Weighted mid-infrared spectral diagnostics \citep{Diaz-Santos2017} indicate that the fraction of the bolometric luminosity contributed by an AGN is less than 15\% in these four objects.
These combined constraints indicate tension in the implied contribution of BH accretion to the total luminosity of the system.
A summary of the source properties and the \nustar observations is given in Table~\ref{table:sample}.

\subsubsection{The Use of Global versus Resolved (Sub)Millimeter Line Ratios}

The selection of these four sources is based on their global HCN/HCO$+$ line ratio.
It has been noted that resolution effects can in some circumstances result in ``contamination'' of any HCN/HCO$^+$ AGN signature by surrounding starburst activity \citep{Izumi2016}.
This is a result of the general behavior that starbursts have HCN/HCO$^+$ ratios $\sim0.9-1.2$ and many AGN have ratios $\gtrsim2$.
Measurement of the line ratio in increasingly larger apertures around AGN with intrinsically high ratios will be diluted due to the inclusion of regions of star-formation with lower ratios.
This would act to reduce the measured line ratio in larger apertures which include increasing contribution from the starburst, and may explain some instances where coarse resolution measurements of AGN show ``starburst'' line ratios.

In contrast, globally \emph{enhanced} HCN/HCO$^+$ line ratios in starburst galaxies are not straightforward to explain via contamination from extended emission associated with star formation.
As any contamination to these global apertures from a starburst decreases the line ratio (as described above), attempting to correct for the contamination would increase the inferred HCN/HCO$^+$ line ratio.
Global line ratios that are enhanced relative to the values typical for star formation are likely reliable indicators of nuclear enhancements \citep[e.g.,][]{Privon2017a}.
Thus, the use of single-dish spectra to select these objects as having elevated HCN/HCO$^+$ line ratios is unlikely to result in false positives due to contamination from extended star formation.

\subsection{X-Rays}

We employ \nustar X-ray observations from 3--79 keV (Section~\ref{sec:nustarobs}) in order to search for emission associated with an AGN.
This energy range is particularly suited to this task as emission above $\gtrsim10$ keV is less affected by high column densities than emission at lower energies.
We supplement the \nustar observations with archival observations of softer X-rays (Section~\ref{sec:softobs}) to anchor constraints on the contribution of X-ray emission from star formation and any unobscured emission from an AGN.

\subsubsection{\nustar Hard X-ray Observations}
\label{sec:nustarobs}

The four targets were observed by NuSTAR for $\sim20$ ks each (see Table\,\ref{table:sample}), in Cycle 3 (Project 3190, PI: G. C. Privon).
The data obtained by the two \nustar focal plane modules, FPMA and FPMB, were processed using the {\it NuSTAR} Data Analysis Software \textsc{nustardas}\,v1.8 within Heasoft\,v6.24, using the calibration files released on UT 2018 August 18.
The cleaned and calibrated event files were produced using \textsc{nupipeline} following standard guidelines.
For all sources we extracted a spectrum at the position of the counterpart, selecting circular regions of 45\,arcsec radius centered on the position of the source.
For the background spectra we used an annular region centered on the source with inner and outer radii of 90 and 150\,arcsec, respectively.

\begin{deluxetable*}{lccccc}
\tablecaption{Sample Properties, Observations, and Measurements}
\tablehead{\colhead{Quantity} & \colhead{Units} & \colhead{IRAS~F06076--2139} & \colhead{NGC~5104} & \colhead{NGC~6907} & \colhead{NGC~7591}}
\startdata
(1) Redshift & \nodata & 0.0374\phantom{0}     &   0.01860     &    0.01541   &  0.01655   \\
(2) $\log_{10}($\LIR/L$_{\odot})$  & \nodata &   11.59   &   11.09     &   11.03    &  11.05   \\
(3) HCN/HCO$^+$ & \nodata &  $>1.8$    &  $1.7\pm0.5$      &  $1.8\pm0.4$     &   $2.0\pm0.4$  \\
(4) 6.2 $\mu$m PAH EQW & ($\mu$m)  & 0.33     &    0.51    &   0.57    &  0.48   \\
(5) MIR-determined $f_{\rm AGN, bol}$  & \nodata &  $0.11\pm0.01$    &   $0.10\pm0.04$     &   $0.09\pm0.04$    &  $0.09\pm0.02$   \\
(6) {\it NuSTAR} Obs Date & (YYYY--MM--DD)  &  2018--01--29    &    2018--01--12    &  2018--05--13     &   2018--01--09  \\
(7) {\it NuSTAR} exposure & (ks)  &  23.1/23.1    &   19.0/19.3    &  19.0/18.9     & 18.3/17.7    \\
(8) {\it Chandra}/ACIS Obs Date & (YYYY--MM--DD)  &   2012--12--12    & \nodata       &   \nodata    &   2009--07--05  \\
(9) {\it Chandra}/ACIS exposure & (ks)   &   14.8    &     \nodata   &    \nodata   &  4.9   \\
(10) {\it Swift}/XRT Obs Date & (YYYY--MM--DD)  & \nodata      &   Stacked     &   Stacked    &  \nodata   \\
(11) {\it Swift}/XRT exposure & (ks)   &   \nodata    &     10.3   &     19.9  &  \nodata   \\
\hline
(12) L$_{2-10\rm\,keV}$ (obs) & (erg s$^{-1}$) & $1.6_{-0.4}^{+0.1}\times10^{41}$ & $1.5^{+0.7}_{-0.6}\times10^{40}$ & $7.2^{+1.9}_{-1.5}\times10^{39}$ & $9.3^{+4.4}_{-4.3}\times10^{39}$\\
(13) \LHX (obs) & (erg s$^{-1}$) & $1.1_{-0.3}^{+0.1}\times10^{42}$ & $<4.05\times10^{40}$ & $<1.83\times10^{40}$ & $<2.83\times10^{40}$ \\
(14) \LAGN (3--24 keV, intrinsic) & (erg s$^{-1}$) & $3.2_{-1.3}^{+2.2}\times 10^{42}$ & \nodata & \nodata & \nodata \\
(15) N$_{\rm H}$  & (\cmcol) & $6.2^{+2.3}_{-1.6}\times10^{23}$ & \nodata & \nodata & \nodata \\
(16) \LAGN/\LIR (N$_H=10^{24}$ \cmcol) & \nodata & \nodata & $<0.002$ & $<0.001$ & $<0.003$ \\
(17) \LAGN/\LIR (N$_H=10^{25}$ \cmcol) & \nodata & \nodata & $<0.090$ & $<0.059$ & $<0.151$ \\
\enddata
\tablecomments{Redshifts (1) and \LIR (2) values are from \citet{Armus2009}, HCN/HCO$^+$ ratios (3) are for the (1--0) transitions from \citet{Privon2015}, and $f_{\rm AGN,bol}$ (4) values are the bolometric AGN fractions obtained by \citet{Diaz-Santos2017} using a combination of mid-infrared diagnostics.
The lower limit for the HCN/HCO$^+$ ratio of IRAS~F06076--2139 is a $3\sigma$ value.
Observation dates are given (6, 8, 10) if the observation was performed in a single block, otherwise ``Stacked'' indicates data were combined from multiple observations.
Observation times (7, 9, 11) are the total on-source time.
For {\it NuSTAR} the two exposure times refer to FPMA and FPMB, respectively.
(12) and (13) give the determined luminosities based on the {\it Chandra}/ACIS or {\it Swift}/XRT and \nustar observations.
(14) and (15) show the intrinsic 3--24 keV luminosity and the obscuring column for the AGN detected with spectral modeling.
(16) and (17) show the AGN luminosity limits for two different assumed values for the obscuring column.
\LAGN/\LIR limits were determined from the \LHX upper limits assuming the noted X-ray obscuring column and a bolometric correction factor of 13.6 \citep[adapted from][]{Vasudevan2009}.
Upper limits for quantities derived from X-ray observations are $1\sigma$.
}
\label{table:sample}
\end{deluxetable*}

\subsubsection{{\it Chandra}/ACIS and {\it Swift}/XRT Archival Observations}
\label{sec:softobs}

Two of the four sources (\object{IRAS F06076--2139} and \object{NGC 7591}) have archival {\it Chandra} \citep{Weisskopf:2000vn} ACIS \citep{Garmire:2003kx} observations.
The {\it Chandra} ACIS data were reduced following the standard procedures, using \textsc{CIAO} v.4.7.9.
We reprocessed the data using the \textsc{chandra\_repro} task and extracted source spectra using circular regions of 5\,arcsec radius.
For the background we used circular regions of radius 10\,arcsec, selected in regions where no other source was present.
Both objects show point-like sources coincident with the optical counterparts.

The other two sources, \object{NGC 5104} and \object{NGC 6907} were observed several times by the XRT detector \citep{Burrows:2005vn} on board the  {\it Neil Gehrels Swift Observatory} \citep{Gehrels:2004kx}.
We used all the data available (for a total of 10.3\,ks and 19.9\,ks, respectively), processing it using the \textsc{xrtpipeline} within \textsc{heasoft\,v6.24} following the standard guidelines.

\section{X-Ray Properties}
\label{sec:xrays}

\subsection{Individual Sources}

\begin{figure*}
\includegraphics[width=\textwidth]{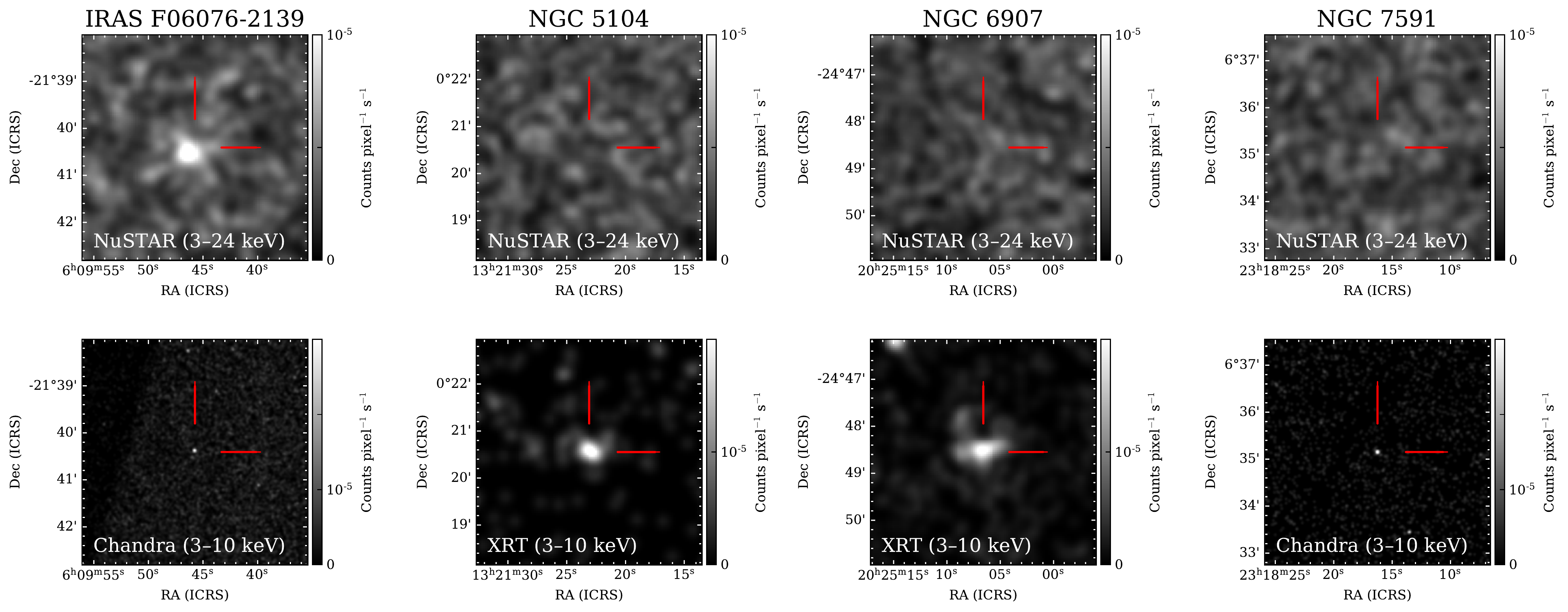}\\
\includegraphics[width=\textwidth]{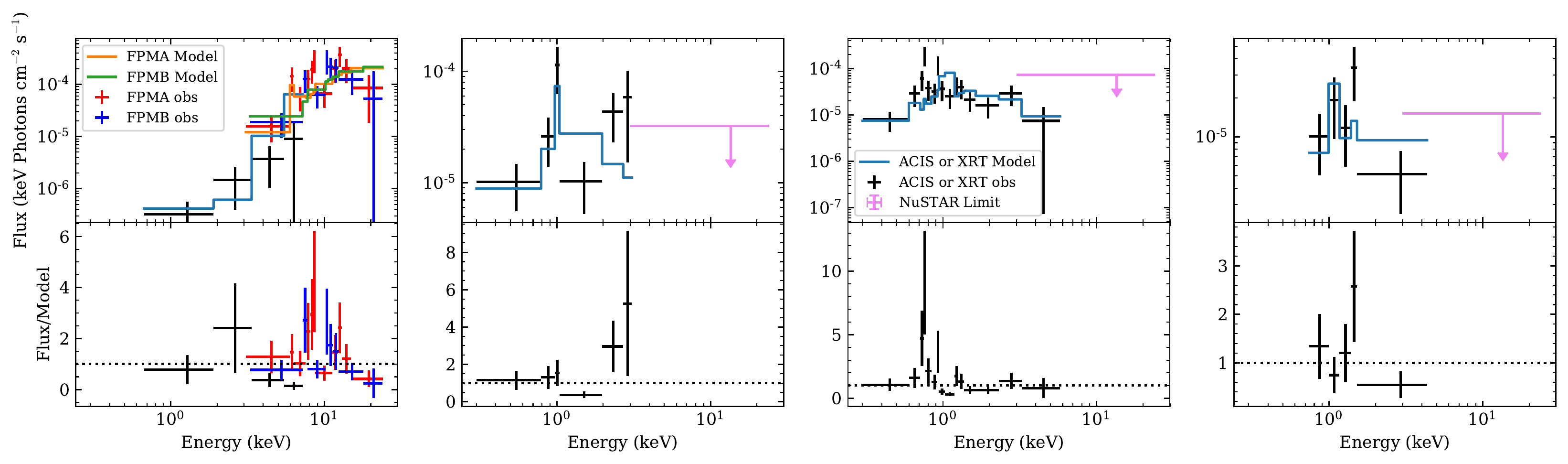}\\
\includegraphics[width=\textwidth]{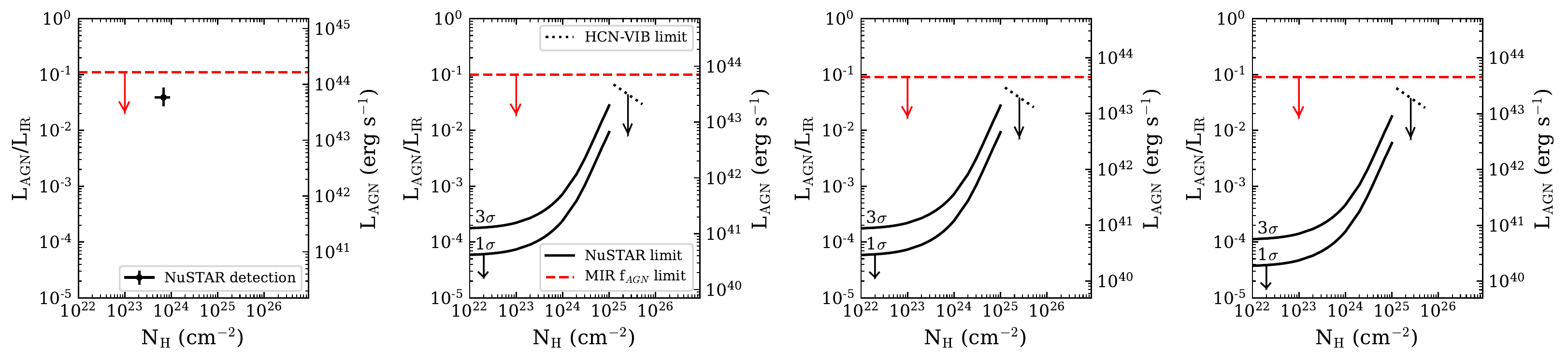}
\caption{A summary of the X-ray observations and modeling results.
Top row: \nustar images of the field.
Second row: \chandra/ACIS or {\it Swift}/XRT images.
Third row: Broadband X-ray spectrum (crosses) with models (solid lines).
Fourth row: Joint constraints on L$_{\rm AGN}$/\LIR and \NH, based on the \nustar observations and models noted in the text.
In the first and second rows, the locations of the target galaxies are indicated by the red lines.
The grayscale images are Gaussian smoothed with a kernel size of 3 pixels and the intensity is a linear stretch.
In the fourth row curves show the 1 and 3$\sigma$ upper limits for L$_{\rm AGN}$/\LIR as a function of column density, except for IRAS~F06075--2139, where we detected an AGN and the point denotes its measured luminosity and obscuring column.
Regions above the curves are excluded by the \nustar data.
The red dashed line marks the upper limits for the bolometric AGN fraction determined from MIR observations \citep{Diaz-Santos2017}.
Finally, the dotted lines show the
$3\sigma$ \LAGN limit computed from HCN-VIB upper limits presented by D. Jeff et~al. (\emph{2020, in preparation}), using the models of \citet{Gonzalez-Alfonso2019}.
}
\label{fig:Xray}
\end{figure*}

Of the four sources in the sample, only IRAS~F06076--2139 was detected in hard X-rays ($\geq 10$\,keV).
The other sources were not detected above the background emission.
In Figure~\ref{fig:Xray} we show the \nustar images, \chandra/ACIS or {\it Swift}/XRT images, X-ray spectra, and the constraints on \LAGN and \NH.
Below we briefly discuss the individual systems; their properties are summarized in Table~\ref{table:sample}.

\paragraph{IRAS~F06076--2139}

The source was detected by our 20\,ks {\it NuSTAR} observation in both the $3-10$ and $10-24$ keV band and both the North and South nuclei were detected by an archival {\it Chandra} observation.
IRAS~F06076--2139 is a double system and the northern component is the dominant source in the far infrared and soft X-rays \citep{Torres-Alba:2018rt}.
It is probable that the IRAM 30m HCN/HCO$^+$ measurement of \citet{Privon2015} is dominated by this northern component, as ALMA observations of the 4--3 HCN/HCO$^+$ transition find emission only in the northern galaxy (D. Jeff et~al. \emph{2020, in preparation}).

We combined the {\it Chandra} spectrum of the South nucleus, which was the one suspected to host an AGN by \cite{Torres-Alba:2018rt}, with the {\it NuSTAR} spectra.
The $0.3-24$\,keV spectrum was first fit with a simple model consisting of a powerlaw [\textsc{tbabs$\times$(zpo)} in XSPEC].
We included Galactic absorption using the \textsc{tbabs} model \citep{Wilms:2000vn}, fixing the column density \NH to the value reported by \citet{Kalberla:2005fk} at the coordinates of the source.
The model results in a good fit, C-stat=222 for 209 degrees of freedom (DOF), and in a flat power-law component ($\Gamma=-0.22_{-0.26}^{+0.25}$).
We measure observed $ L_{3-24\rm\,keV}=1.1_{-0.3}^{+0.1}\times10^{42}\rm\,erg\,s^{-1}$, and $ L_{2-10\rm\,keV}=1.6_{-0.4}^{+0.1}\times10^{41}\rm\,erg\,s^{-1}$.
The 2--10\,keV luminosity from the North nucleus is higher ($L_{2-10\rm\,keV}=2.3_{-0.4}^{+0.2}\times10^{41}\rm\,erg\,s^{-1}$) and softer, and the {\it Chandra} spectrum can be represented by a thermal plasma component (\textsc{apec}).

We compute the expected X-ray luminosity due to star formation by starting with the total IR luminosity of the system [$\log_{10}($\LIR/\Lsun$)=11.59$] and computing the SFR following Eq.\,4 in \citet{Murphy2011}, SFR$=58\rm\,M_{\odot}\,yr^{-1}$.
We then use the \cite{Lehmer2016} relationship to calculate the 2--10\,keV luminosity due to SF, $L_{2-10\rm\,keV}^{\rm SF}=2.7\times 10^{41}\rm\,erg\,s^{-1}$.
This value is only slightly lower than the sum of the observed 2--10\,keV luminosities of the two nuclei, which implies that most of the emission below 10\,keV could arise from star formation.
However, the flat power-law component seen in the Southern nucleus, together with the clear detection above 10\,keV by {\it NuSTAR}, argue for the presence of an obscured AGN.
We therefore fit the spectra using the X-ray torus model of \citet{Balokovic18aa} [\textsc{tbabs$\times$(zcut+atable\{borus\}+ztbabs$\times$cabs$\times$zcut)}], fixing $\Gamma=1.8$ and the cutoff energy to $E_{\rm cut}=300$\,keV both in the torus model and in the secondary powerlaw (which takes into account both contribution from X-ray binaries and Thomson scattered X-ray radiation).
The model yields a good fit (C-Stat/DOF=202/208) and the column density of the absorber we obtained is $6.2^{+2.3}_{-1.6}\times10^{23} \rm\,cm^{-2}$, while the intrinsic luminosity of the AGN in the 3--24\,keV band is $3.2_{-1.3}^{+2.2}\times 10^{42}\rm\,erg\,s^{-1}$.
Using a bolometric correction of $13.6$ \citep[derived from ][]{Vasudevan2009} the X-ray luminosity (corrected for both intrinsic and Galactic absorption) implies \LAGN/\LIR$\approx3\%$.

We also tested whether a reflection-dominated model could reproduce the X-ray spectrum of the source, by setting the column density of the torus model to $N_{\rm H}=10^{25}\rm\,cm^{-2}$.
We obtained a significantly worse statistic than considering a transmitted-dominated model both keeping the photon index fixed to $\Gamma=1.8$ (C-stat/DOF=315/209) and leaving it free to vary (C-stat/DOF=311/208).

While our {\it NuSTAR} observations do not resolve the pair, it is probable that the obscured AGN is associated with the southern nucleus.
Thus the AGN, contributing a small fraction of the total luminosity, may be unassociated with the region showing enhanced HCN emission.
For the sake of completeness we also performed a spectral fit using the {\it NuSTAR} data and only the {\it Chandra} counts associated with the northern nucleus, and we find consistent AGN properties (e.g., \NH$=6.8_{-2.4}^{+2.7}\times10^{23}\rm\,cm^{-2}$).
For this fit we added a thermal plasma component (\textsc{apec}) to reproduce the soft emission.
We can therefore conclude that the properties of the AGN in IRAS~F06076--2139 do not depend strongly on the assumed location of the AGN.
ALMA Band 7 observations of the 4--3 transitions of HCN and HCO$^+$ localize it to the northern galaxy (D. Jeff et~al. \emph{2020, in preparation}).

\paragraph{NGC~5104}

This source was not detected by \nustar, and we determined a $3\sigma$ upper limit to the observed $3-24$ keV luminosity of $<4.05\times10^{40}$ erg s$^{-1}$.
NGC~5104 was clearly detected by {\it Swift}/XRT, and the spectrum is well fit by a single thermal plasma model (\textsc{tbabs$\times$apec} in XSPEC), resulting in C-Stat/DOF=26/24.
This spectral model is consistent with the idea that most of the X-ray emission in this object is related to star formation.
The observed 2$-$10\,keV luminosity of this object is $1.5_{-0.6}^{+0.7}\times10^{40}\rm\,erg\,s^{-1}$.
Following the same approach used for IRAS~F06076--2139, we find that the predicted 2--10\,keV luminosity associated with the SFR is $9.2\times 10^{40}\rm\,erg\,s^{-1}$, which is even higher than the observed luminosity, suggesting that all the X-ray emission comes from star formation in this object.

\paragraph{NGC~6907}

NGC~6907 was not detected by {\it NuSTAR}, and we determined a $3\sigma$ upper limit to the observed $3-24$ keV luminosity $<1.83\times10^{40}$ erg s$^{-1}$.
The source is clearly detected by {\it Swift}/XRT, and the spectrum can be well fit (C-Stat/DOF=67/58) with the same model adopted for NGC~5104.
The observed X-ray luminosity in the 2--10\,keV range is $7.3_{-1.5}^{+1.9}\times10^{39}\rm\,erg\,s^{-1}$, which is lower than that expected by star formation ($L_{2-10}^{\rm SF}=8.4\times 10^{40}\rm\,erg\,s^{-1}$), suggesting that there is no significant AGN contribution to the X-ray spectrum.

\paragraph{NGC~7591}

NGC~7591 was not detected by \nustar.
We determined a $3\sigma$ upper limit to the observed $3-24$ keV luminosity of $<2.83\times10^{40}$ erg s$^{-1}$.
The soft X-ray spectrum from this source is well fit by a thermal plasma model (\textsc{tbabs$\times$apec}), resulting in C-Stat/DOF=18/19.
The X-ray luminosity is $L_{2-10\rm\,keV}=9.2_{-4.3}^{+4.4}\times10^{39}\rm\,erg\,s^{-1}$, which is also lower than that predicted from star formation ($L_{2-10}^{\rm SF}=8.3\times 10^{40}\rm\,erg\,s^{-1}$), implying that the X-ray emission of this source is dominated by star formation.

\subsection{Do HCN Enhancements Indicate Buried AGN?}

Here we summarize the \nustar search for AGN in these systems with elevated HCN/HCO$^+$ ratios (Section~\ref{sec:hx-search}).
We also briefly discuss ALMA observations of these sources, which we use to further constrain heavily obscured nuclear activity in these systems (Section~\ref{sec:HCNVIB}).

\subsubsection{Hard X-Ray Search}
\label{sec:hx-search}

Out of the four sources, an indication of an obscured AGN is only present for IRAS~F06076--2139.
However, the AGN contributes a minor fraction (3\%) to the total luminosity of the system and is likely not spatially associated with the molecular gas.

The \nustar upper limits enable us to place joint constraints on \LAGN and \NH for the other three sources.
We take the \LHX upper limits and apply corrections for a range of obscuring columns determined from the Mytorus models of \citet{Murphy2009}.
We then convert this to an upper limit on \LAGN (for that assumed \NH) using a \LHX bolometric correction of 13.6 derived from \citet{Vasudevan2009}.
In the bottom row of Figure~\ref{fig:Xray} we show the joint constraints on \NH and \LAGN based on this approach.
These \nustar observations indicate that AGN cannot contribute more than $\sim10\%$ unless the obscuration is $>10^{25}$\cmcol.

In the bottom row of Figure~\ref{fig:Xray} we also show the average bolometric AGN fractional contributions, $f_\mathrm{AGN}$, computed by performing a survival analysis on the individual AGN fractions derived from both atomic and dust MIR tracers of AGN activity \citet{Diaz-Santos2017}.
In the case of these four objects, none had \nev detections.
Two sources (NGC~5104 and NGC~6907) had \oiv detections, though the inferred AGN fractions are consistent with no contribution of the AGN to the mid-infrared luminosity.
The PAH EQW \citep[see Table~\ref{table:sample} and][]{Stierwalt2013} was also included in the AGN fractions computed by \citet{Diaz-Santos2017}.
The AGN fraction derived from the PAH EQW requires the adoption of a baseline value for a source free of SMBH accretion.
In our original identification of these sources as pure starburst or starburst-dominated \citep{Privon2015}, we adopted the pure \emph{starburst} value from \citet{Brandl2006}.
In contrast, when computing the AGN fractions, \citet{Diaz-Santos2017} assumed a higher PAH EQW reference value typical of normal star-forming galaxies, following \citet{Stierwalt2013} and \citet{Veilleux2009}.
The baseline EQW value for a pure starburst is lower than that of a normal star-forming galaxy because in the former, a larger fraction of the starlight is reprocessed by dust and subsequently emitted as infrared continuum, resulting in brighter continuum and lower EQW.

As the \citet{Diaz-Santos2017} mean AGN fractions were computed using a higher PAH EQW than is seen in pure starbursts, we consider the $f_\mathrm{AGN}$ values in Table~\ref{table:sample} and Figure~\ref{fig:Xray} as \emph{upper limits} to the true AGN fraction for these systems.
We defer a detailed examination of the agreement between hard X-ray and other multiwavelength AGN indicators (including those in the MIR) to C. Ricci (et al. 2020 \emph{in preparation}),
but later briefly comment on scenarios varying the distribution of gas and dust, and the impact on these AGN diagnostics (Section~\ref{sec:nodust}).
If these MIR indicators are accurate, they would imply the \nustar observation would be consistent with \NH$>10^{25}$ \cmcol.
However, such column densities imply that the MIR would be optically thick, and therefore diagnostics that use spectral features in this range may not be accurate.
In addition, modeling of dust obscuration of starburst and AGN energy sources suggests that modest unobscured lines of sight to a starburst can dilute mid-infrared signatures of AGN \citep{Marshall2018}.

\subsubsection{Vibrationally Excited HCN}
\label{sec:HCNVIB}

One possible explanation for the lack of detected hard X-ray emission in the three sources is extremely high column densities, in excess of $\sim10^{25}$ \cmcol.
Such high levels have been reported for other systems \citep{Sakamoto2013,Scoville2015,Aalto2019} and it would be extremely difficult to see X-rays from a sufficiently buried AGN, particularly if the covering factor is $\sim1$.
However, a sufficiently luminous, buried energy source may radiatively pump HCN molecules, leading to HCN-VIB emission \citep[e.g.,][]{Ziurys1986,Aalto2015a}.
Thus, if there are energetically important AGN with sufficiently high obscuration that we do not see it with \nustar, we may expect HCN-VIB emission in the (sub)millimeter.

Limits on the HCN-VIB (4--3) emission will be presented by D. Jeff et~al. (\emph{2020, in preparation}), based on ALMA Band 7 observations of these four sources.
We use those HCN-VIB upper limits in conjunction with Figure~13 of \citet{Gonzalez-Alfonso2019} to determine upper limits for \LAGN/\LIR as a function of column density
\footnote{HCN-VIB/\LIR has a factor of a few dependence on the IR luminosity surface density, $\Sigma_\mathrm{IR}$, but we conservatively use, for a given \NH, the lowest HCN-VIB/\LIR value for all relevant $\Sigma_\mathrm{IR}$.
Any $\Sigma_\mathrm{IR}$ variations would \emph{increase} the predicted HCN-VIB luminosity over what we show here.}.
These constraints are shown as the dotted lines in the bottom row of Figure~\ref{fig:Xray} and suggest that any AGN with N$_\mathrm{H}>10^{25}$ \cmcol would likely contribute less than $10\%$ of the bolometric luminosity to each system.

\subsection{The Emergent X-Ray Flux and Dust-free Gas Obscuration}

Here we use the X-ray upper limits to investigate the viability of X-ray driven chemistry in explaining the observed HCN enhancements (Section~\ref{sec:xraychem}).
We also briefly describe a scenario in which X-rays from an AGN are attenuated by dust-free gas (i.e., within the dust sublimation radius), and the implications of this scenario for multi-wavelength AGN indicators (Section~\ref{sec:nodust}).

\subsubsection{The Viability of X-Ray Driven Chemistry in Explaining HCN enhancements}
\label{sec:xraychem}

If the column density is high enough to effectively prevent any X-rays from escaping, they may also be unable to penetrate sufficiently deeply into the molecular gas to drive changes in the chemistry.
Based on the \nustar curves of Figure~\ref{fig:Xray}, we tentatively rule out the existence of AGN contributing more than $\sim10\%$ of \Lbol in NGC~5104, NGC~6907, and NGC~7591, even if the AGN is heavily obscured.
The \citet{Meijerink2007} XDR models indicate that HCN/HCO$^+>1$ due to X-ray driven chemistry is expected only when $F_X>10$ erg s$^{-1}$ cm$^{-2}$ (incident flux on the cloud surface, over the energy range of 1--100 keV, where they assume a powerlaw energy spectrum of $F(E)\propto E^{-0.9}$).
We use the observed upper limits on the hard X-ray flux from the AGN to place limits on the X-ray flux at the location of the circumnuclear material where the HCN and HCO$^+$ line emission is expected to arise.
Taking our $3-24$ keV limits and correcting to the $1-100$ keV energy range (assuming the same $\alpha=0.9$) we estimate the flux experienced by molecular clouds in the nuclear regions of these systems is at least a factor of 10 below what is predicted to lead to HCN enhancements.
This suggests that the X-ray flux currently incident on circumnuclear clouds is unlikely to be sufficient to drive the chemistry necessary to explain the HCN enhancements.
In Section~\ref{sec:timescale} we discuss the timescales of significant AGN variability and the chemical evolution of the gas.

\subsubsection{X-ray Obscuration Within the Dust Sublimation Radius}
\label{sec:nodust}

For a scenario in which there is an AGN, the mid-infrared and hard X-ray discrepancy may be resolved if the X-rays are obscured by gas that is largely within the dust sublimation radius.
In this case, the small-scale dust-free gas would obscure the X-rays, but the dust on several parsec scales and larger would feel the influence of the reprocessed optical/UV radiation.
This picture is potentially supported by observations which find that optical extinctions underestimate the X-ray obscuring columns toward AGN \citep[e.g.,][]{AlonsoHerrero1997,Merloni2014,Burtscher2016}, though as those authors discuss, other scenarios can also explain the data.

The \nustar upper limits for the three nondetections indicate that values of \Lbol$\gtrsim 10^{43}$ erg s$^{-1}$ would require \NH$\gtrsim 10^{25}$ \cmcol (Figure~\ref{fig:Xray}).
This bolometric luminosity corresponds to a dust sublimation radius of 0.04 pc \citep[see][]{Nenkova2008}.
Assuming a spherically symmetric gas configuration with \NH$=10^{25}$ \cmcol contained within the dust sublimation radius suggests a mean density of $\approx8\times10^{7}$ \cmc.

A consequence of this scenario is that the heavy attenuation of the X-rays on small size scales would ensure that no significant X-ray flux would be seen by the molecular clouds in the nuclear regions (e.g., Section~\ref{sec:xraychem}).
Thus the X-rays from an AGN with such a high dust-free obscuring column could play no role in driving the chemistry of molecular clouds or influencing the HCN/HCO$^+$ line ratio.

We ran a radiative transfer experiment with CLOUDY \citep{Ferland2017} of a \Lbol=10$^{43}$ erg s$^{-1}$ luminosity AGN embedded in a uniform, dust-free, solar metallicity gas cloud with a total column density of \NH$=10^{25}$ \cmcol and an outer size of 0.04 pc.
Effectively, no ionizing photons ($\mathrm{E}>13.6$~eV) escape this obscuration.
This lack of ionizing photons would likely prevent MIR indicators such as \nev and \oiv from being present.
We thus conclude that the scenario of dust-free obscuring columns is not a viable mechanism for explaining any apparent tension between the MIR and X-ray results for these systems.

\section{The Link Between HCN Emission and AGN}
\label{sec:results}

\begin{figure*}
    \centering
\includegraphics[width=\columnwidth]{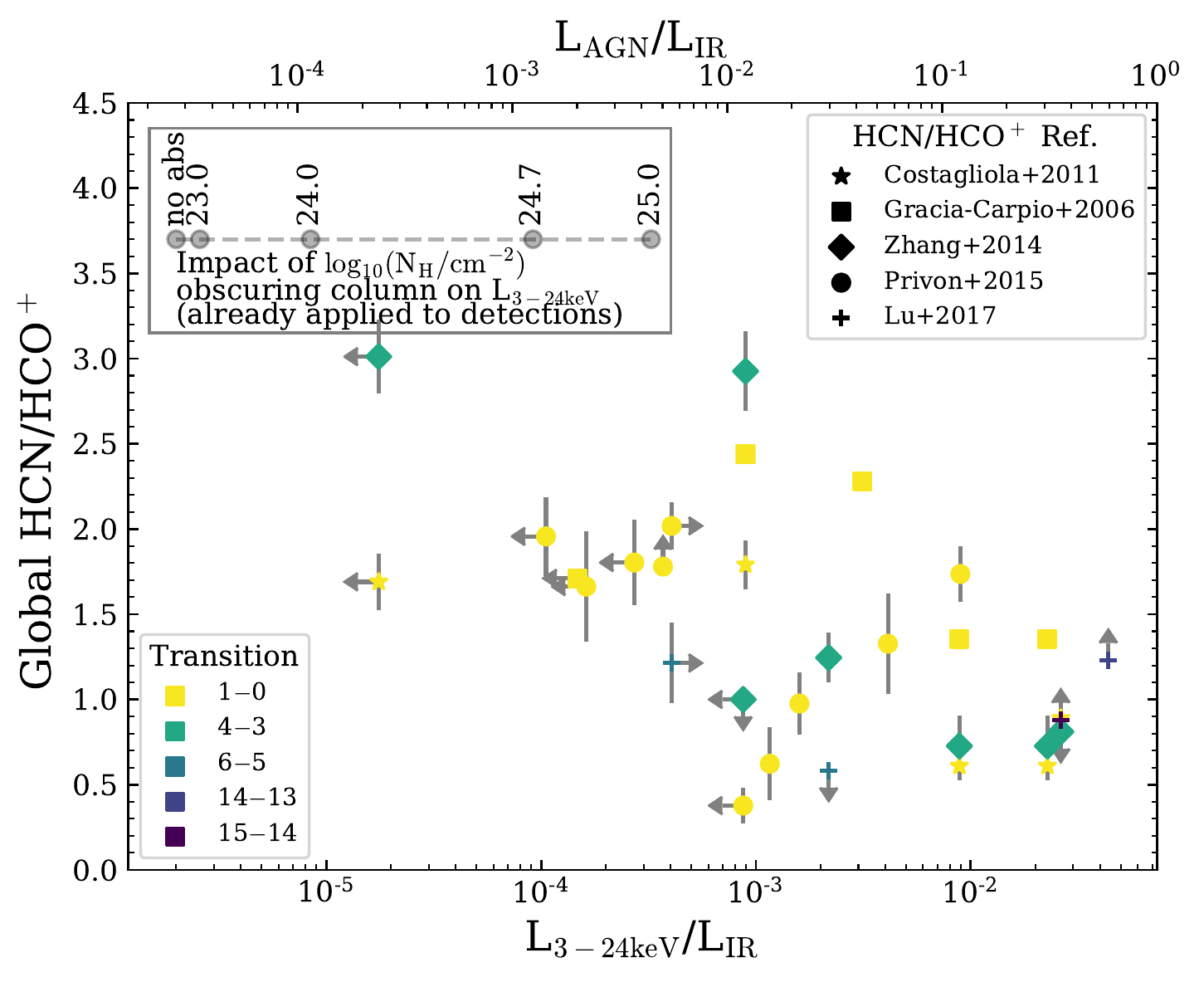}
\includegraphics[width=\columnwidth]{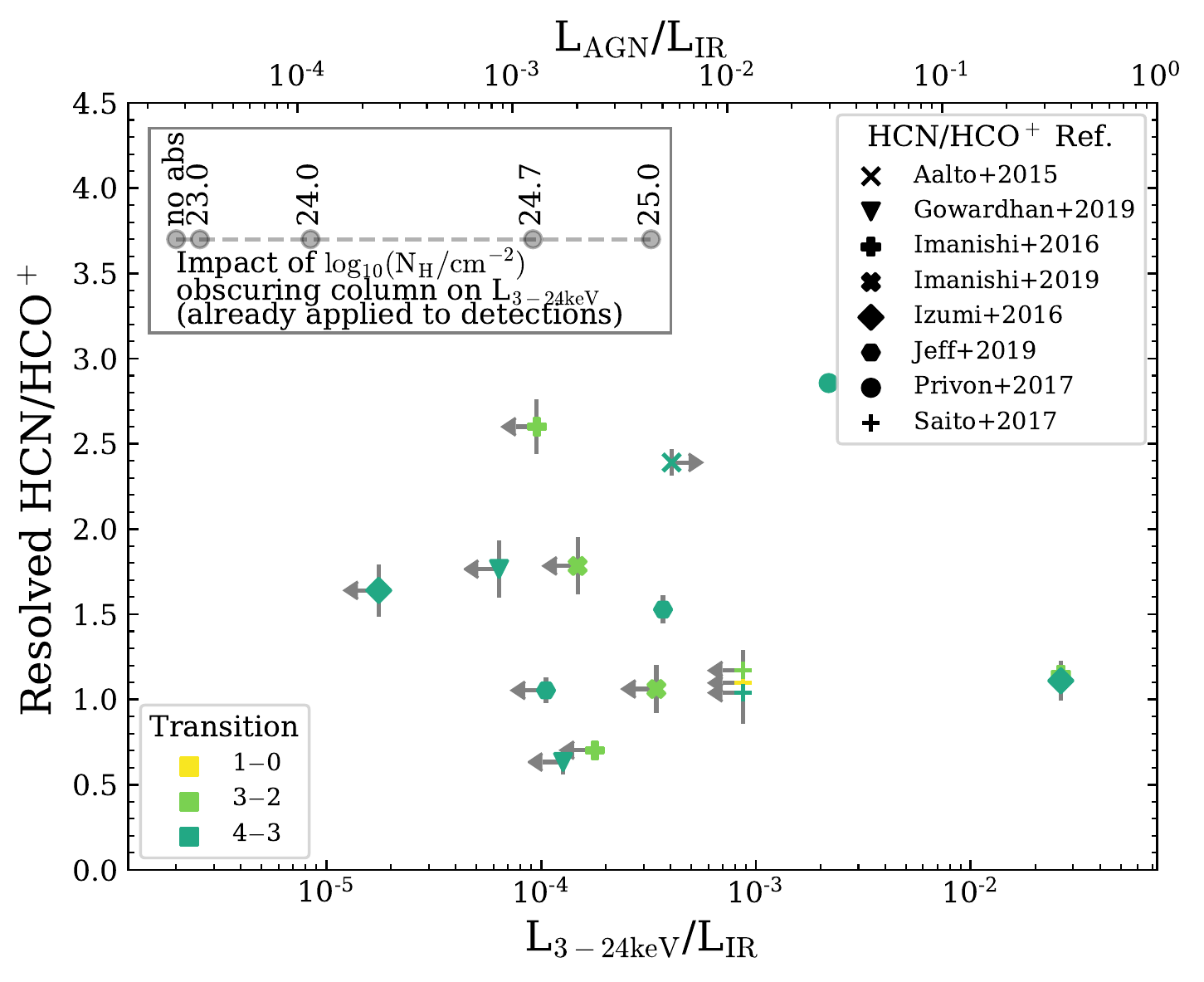}
\caption{A comparison of \LHX/\LIR with various HCN/HCO$^+$ line ratios for unresolved (left) and resolved (right) molecular observations.
    The \LHX values for detections are corrected for both the MW foreground absorption as well as the intrinsic AGN absorption (where the latter is determined from spectral fitting of the X-rays; C. Ricci et al. 2020 \emph{in preparation}).
The \LHX upper and lower limits have not been corrected for any starburst contribution or obscuration correction.
The gray points and dashed line in the upper left of each panel show the magnitude of absorption correction to transform observed $3-24$ keV luminosities into intrinsic values.
This correction has already been applied to the X-ray detections, but indicates how the upper and lower \LHX/\LIR limits may move toward the right under an (unknown) absorption correction.
The top x-axis shows L$_{AGN}$/\LIR, assuming the X-rays originate from an AGN and use a bolometric correction of 13.6 \citep[adapted from][]{Vasudevan2009}.
Note that the maximum of the x-axis of both plots corresponds to \LAGN=\LIR, for this assumed bolometric correction.
In both panels the colors denote ratios of HCN/HCO$^+$ with
$J_{\rm upper}=1$ (\crule[0.97,0.90,0.13]{6pt}{6pt}),
3 (\crule[0.48,0.82,0.31]{6pt}{6pt}),
4 (\crule[0.13,0.66,0.51]{6pt}{6pt}),
6 (\crule[0.16,0.47,0.56]{6pt}{6pt}),
14 (\crule[0.26,0.27,0.53]{6pt}{6pt}),
or 15 (\crule[0.27,0.00,0.33]{6pt}{6pt}).
Note that some galaxies may appear multiple times in either if the molecular emission has been measured from more than one $J_{\rm upper}$ level.
There is no apparent correlation between the global or resolved HCN/HCO$^+$ ratios and the \LHX/\LIR ratio (e.g., Figure~\ref{fig:statistics}).
Literature data were taken from \citet{GraciaCarpio2006,Baan2008,Imanishi2009,Costagliola2011,Privon2015,Aalto2015a,Imanishi2016b,Imanishi2016c,Izumi2016,Zhang2016,Saito2017,Lu2017,Imanishi2019}; A. Gowardhan et~al. (2020, \emph{in preparation}); and D. L. Jeff et~al. (2020, \emph{in preparation}).
}
\label{fig:ratiocomp}
\end{figure*}

We now turn to the broader question of the link between the HCN/HCO$^+$ ratio and the presence/strength of the AGN.
For this study we employ the \nustar observations of GOALS galaxies from \citet{Teng2015,Ricci2017} and C. Ricci et al. (2020; \emph{in preparation}).
In total 54 GOALS systems have been observed with \nustar, and many have literature measurements for the HCN/HCO$^+$ ratio.
We combine the \LHX measurements and upper limits with literature measurements of the galaxy-integrated and spatially resolved HCN and HCO$^+$ transitions, independent of any previously published evidence of AGN activity.
To examine any relationship between these (sub)millimeter line ratios and the bolometric AGN fraction (as diagnosed by the hard X-rays), in Figure~\ref{fig:ratiocomp} we show \LHX/\LIR as a function of the HCN/HCO$^+$ luminosity ratio (computed in L$^{\prime}$ units of K km s$^{-1}$ pc$^2$).
The \LHX detections have been corrected for absorption by the MW as well as intrinsic AGN absorption (the latter determined from fits to the X-ray spectrum; C. Ricci et al. 2020 \emph{in preparation}), upper and lower limits cannot be corrected for the unknown absorption.
The same figure also shows \LAGN/\LIR where we have assumed a bolometric correction from \LHX to \LAGN of 13.6 \citep[adapted from][]{Vasudevan2009}.

\begin{figure*}
\includegraphics[width=\textwidth]{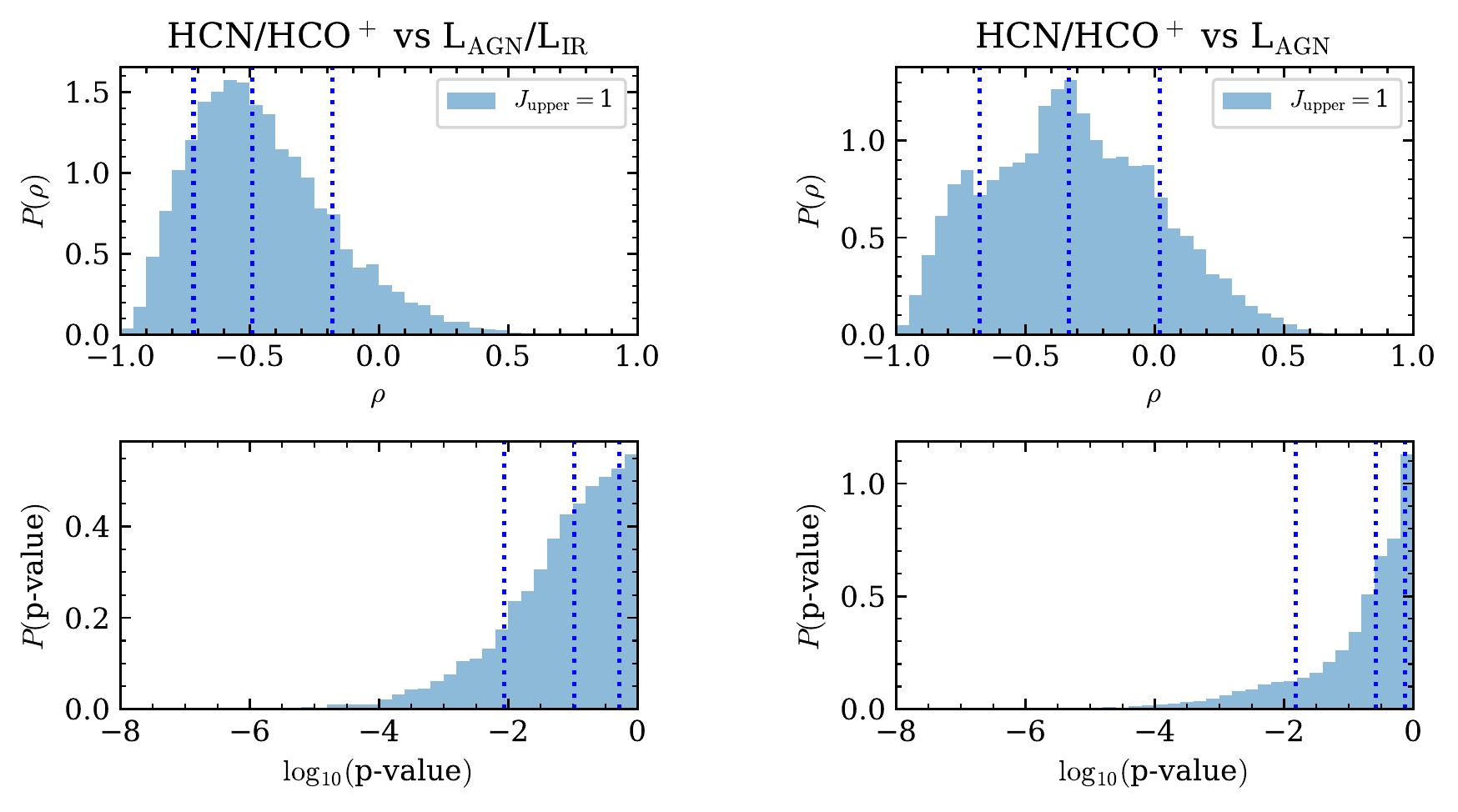}
\caption{The probability density distribution of the \citet{Spearman1904} rank correlation coefficient, $\rho$ (top), and the $p$-value (bottom) between the HCN/HCO$^+$ (1--0) line ratio and the L$_\mathrm{AGN}$/\LIR ratio (left) and between the HCN/HCO$^+$ (1--0) line ratio and L$_\mathrm{AGN}$ (right).
The PDFs were computed using bootstrapping and perturbation following \citet{Curran2014}.
Vertical lines in each panel denote the 16, 50, and 84 percentiles (from left to right; see text for values).
Based on the $p$-values we cannot rule out the null hypothesis that the HCN/HCO$^+$ (1--0) ratio is uncorrelated with either L$_\mathrm{AGN}$/\LIR or L$_\mathrm{AGN}$.
There was an insufficient number of detections for other combinations of the HCN and HCO$^+$ lines to perform the analysis for other $J_{\rm upper}$ levels or for resolved measurements.
}
\label{fig:statistics}
\end{figure*}

We perform a correlation analysis of the global HCN/HCO$^+$ (1--0) ratio and the L$_\mathrm{AGN}$/\LIR ratio\footnote{There are insufficient measurements in other transition pairs or for resolved measurements to perform the same analysis.}.
We compute the \citet{Spearman1904} rank correlation coefficient using a Monte Carlo approach following \citet{Curran2014} as implemented in \texttt{pymcspearman}\footnote{\url{https://github.com/privong/pyMCspearman/}}.
The resulting probability distribution for $\rho$ and the $p$-value are shown in Figure~\ref{fig:statistics}.
We find $\rho=-0.48^{+0.30}_{-0.23}$ with a $p$-value$ = 0.11^{+0.41}_{-0.10}$, where the values quoted are the median and range to the 16/84 percentiles.
Based on the $p$-values we cannot reject the null hypothesis that the HCN/HCO$^+$ luminosity ratio and the L$_\mathrm{AGN}$/\LIR (i.e., AGN fraction) are uncorrelated.
We perform the same analysis on the HCN/HCO$^+$ (1--0) ratio and L$ _\mathrm{AGN}$ itself, also finding we cannot reject the null hypothesis that the two quantities are uncorrelated ($\rho=-0.34\pm0.35$, $p\mathrm{-value}=0.26^{+0.48}_{-0.25}$).
We find there is no evidence that the millimeter line ratio is correlated with the AGN fraction or the present-day AGN luminosity.
We thus conclude that the HCN/HCO$^+$ ratio does \emph{not} trace the fractional AGN contribution to the total galaxy luminosity or the overall AGN luminosity itself.

\subsection{Timescale Considerations}
\label{sec:timescale}

One possible complication in assessing if HCN enhancements trace AGN is whether the timescale for significant AGN variability differs from the timescale of the physical process that leads to the HCN enhancement.
AGN are known to exhibit significant variability on many timescales \citep[e.g.,][]{Hickox2014,Schawinski2015,Sartori2018b}, including up to $10^4-10^6$ years
\footnote{The amplitudes and timescales for AGN variability depend significantly on the observed wave band, but here we focus our discussion on bolometric variability.
    In particular we are interested in variability with large amplitudes (e.g., $\gtrsim 1$ dex) over long timescales (both comparable to the chemical timescale and/or the light crossing time for the molecular region emitting the lines of interest).
Variability that does not meet these requirements is unlikely to result in coherent changes in the molecular line ratios.}.
Such variability would affect the energetic input into the surrounding interstellar medium.
The enhancement of HCN due to an AGN has been proposed as being due to 1) X-ray driven chemistry, 2) cosmic ray driven chemistry, 3) mechanical heating of molecular clouds, or 4) IR pumping of HCN into its $v_2=1$ state.

\subsubsection{Chemical Evolution Timescales}
\label{sec:chemevol}

We argue in this work that there is no evidence that HCN enhancements are convincingly linked to presently observed AGN activity (diagnosed by the hard X-rays).
However, it is possible that the AGN can have an impact on the molecular gas properties, without molecular tracers being useful diagnostics of whether SMBH accretion is present.
If the timescale for any of the above physical processes is significantly different from the AGN turn on or shutdown timescales, the HCN enhancements would not be a good tracer of AGN.
Thus, assessing the timescales for the above physical processes is relevant to understanding the physical origin of HCN enhancements (despite the fact that these enhancements do not appear to be robust in identifying presently accreting SMBHs).
Looking toward X-ray and CR-driven chemistry, \citet{Viti2017} studied the time-dependent chemical evolution of molecular gas after an enhancement of cosmic-ray flux, meant to mimic the impact of cosmic rays and/or X-rays from an AGN.
Their chemical model consisted of two phases, where Phase I simulated the formation of molecular dense gas under galactic conditions, while in Phase II the cosmic-ray flux was varied from the galactic value to several factors higher.
\citet{Viti2017} showed that, if the cosmic-ray ionization rate is not enhanced compared to the galactic one, even with constant cosmic-ray ionization rate, the simulated molecular clouds had not reached a chemical steady state after 1 Myr.

\begin{figure}
    \includegraphics[width=\columnwidth]{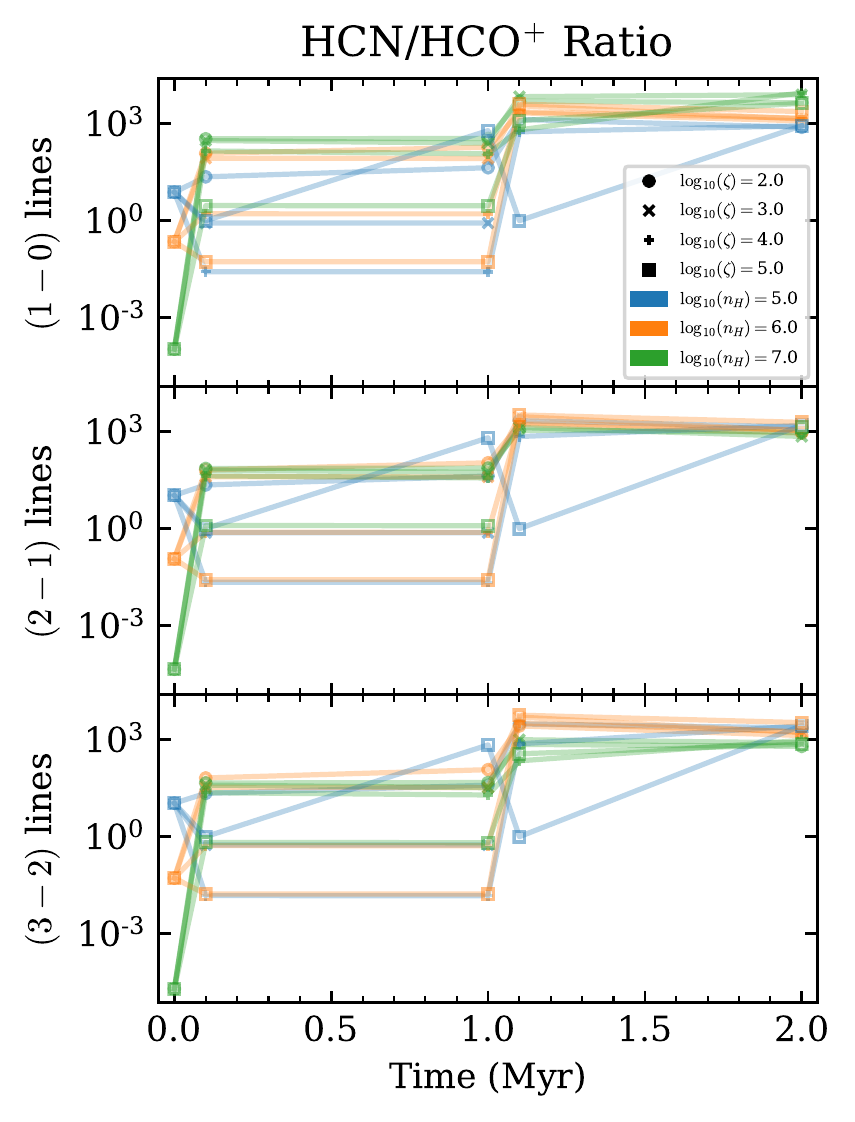}
    \caption{The evolution of HCN/HCO$^+$ line ratios for several $J_\mathrm{upper}$ transitions as a function of time in a chemical simulation.
        Symbols denote different cosmic-ray ionization rates ($\zeta$, normalized to the galactic cosmic-ray flux) during Phase II ($0-1$ Myr) while colors denote different initial densities ($n_\mathrm{H}$).
        During Phase III ($1-2$ Myr) the cosmic ray flux is reduced back to the Galactic value and the chemical evolution is followed; this phase is intended to approximate the molecular gas chemistry after rapid cessation of SMBH accretion.
        We note that these models are general and it is not appropriate to compare these values with observed intensity ratios; the models do not include any observational considerations, including resolution effects \citep[see e.g.,][]{Viti2017}.
    }
    \label{fig:chemistry}
\end{figure}

To explore the chemical evolution after the cessation of accretion, we extend the modeling of \citet{Viti2017} by adding a Phase III, where the enhanced cosmic ray flux is reduced back to the average Galactic flux and the chemical evolution is followed for an additional 1 Myr (S. Viti et~al. \emph{in preparation}).
For the details of the included physical processes we refer the reader to \citet{Viti2017}.
In Figure~\ref{fig:chemistry} we show the time evolution of the HCN/HCO$^+$ ratio
\footnote{These models are not intended to match specific observations, thus they do not include observational effects.}
for several $J_\mathrm{upper}$ levels as a function of time.
Phase II (enhanced cosmic ray flux) occurs between $0-1$  Myr and Phase III (Galactic cosmic ray flux) is between $1-2$ Myr.
The line ratios show significant evolution in both phases, but we note that during and after Phase III the line ratios do not return to their state prior to the onset of the enhanced cosmic ray flux (i.e., Prior to Phase II).
This lends further support to the conclusion that there is no one-to-one mapping of line ratios with AGN activity.

\subsubsection{Radiative Pumping of HCN}

The last relevant physical mechanism is emission from HCN-VIB.
This requires a warm (T$>100$ K), high column density source; also, the strength of HCN-VIB is likely promoted if HCN has an enhanced abundance.
The radiative timescale for HCN-VIB is short, so the lifetime of HCN-VIB emission is set by the cooling time for the central molecular gas, as the primary consideration is the existence of the $T_b>100$ K radiation field.
We use DESPOTIC \citep{Krumholz2014} to estimate the cooling time for a typical HCN-VIB source.
We include CO, C, O, and H$_2$O as cooling species, with properties drawn from \citet{Schoier2005}.
All species were set to standard abundance values, except H$_2$O, which was set to a relative abundance of $10^{-6}$ based on modeling of Herschel observations in similar sources \citep{Gonzalez-Alfonso2014}.
The CO, ortho, and para-H$_2$O species together dominate the cooling luminosity and all contribute approximately in equal measure ($\Lambda\approx8\times10^{-24}$ erg s$^{-1}$ nucleus$^{-1}$ each).
Adopting fiducial values for the highly obscured CONs \citep[$\mathrm{T}_\mathrm{gas}=\mathrm{T}_\mathrm{dust}=200$ K; $n_\mathrm{H}=10^5$ \cmcol; \NH$=10^{25}$\cmcol;][]{Aalto2019}, we find a cooling time of $\sim100$ yr.
This is considerably shorter than AGN flickering timescale.

For highly obscured systems (\NH$>10^{25}$ \cmcol) the radiative diffusion time scale is $\sim10^{4}$ yr \citep{Gonzalez-Alfonso2019}
and the H$_2$O lines are typically seen in absorption \citep{Gonzalez-Alfonso2012,Gonzalez-Alfonso2013} and thus will not contribute to cooling.
If line cooling is inefficient, then the radiative diffusion time will set the cooling time and it will be $\sim10^{4}$ yr.
Together the cooling time and photon diffusion time suggest HCN-VIB emission (and associated blending with HCN $v=0$ lines) should disappear approximately in step with SMBH accretion (if accretion is the dominant energy source).

Though the HCN-VIB emission would likely disappear on timescales similar to that of $>1$ dex AGN flickering, the chemical evolution may be much slower.
Thus, HCN enhancements can persist long past the time in which an AGN has turned off, consistent with our finding that HCN enhancements are poor indicators of current SMBH growth.

\section{Conclusions}
\label{sec:conclusions}

We present \nustar hard X-ray observations of four MIR-classified starbursts with HCN/HCO$^+$ ratios that are elevated relative to values typically associated with star formation.
These observations directly test the proposed use of HCN as a tracer of embedded black hole growth.
We find:

\begin{itemize}
    \item{Three of the four \nustar targets are not detected in hard X-rays, suggesting \LAGN/\LBOL less than $0.3-1.2\%$ ($3\sigma$) if any AGN is Compton thick (N$_{\rm H}\geq 10^{24}$ \cmcol).
        The limits on the AGN fraction increase to a few tens of percent if the column densities are $10^{25}$ \cmcol or greater.
        If such high columns and high AGN fractions are present, we may expect the objects to show (sub)millimeter emission from vibrationally excited HCN; however, emission from HCN-VIB has not been detected in these objects.
The fourth target, IRAS~F06076--2139, was detected with \nustar and the hard X-ray excess suggests the presence of an AGN contributing $\sim3\%$ of the bolometric luminosity, with an obscuring column of $\simeq6.8\times10^{23}$ \cmcol.
However, the AGN in IRAS~F06076--2139 may not be spatially associated with the observed HCN enhancement.}
    \item{In a literature compilation of millimeter and X-ray measurements there is no evidence of a correlation between the HCN/HCO$^+$ (1--0) ratio and either the AGN fraction (L$_{\rm AGN}$/\LIR) or the AGN luminosity.
    This suggests that even when an AGN is present, the (sub)millimeter line ratios do not reflect the current energetics or relative importance of the AGN.}
\item{Based on X-ray tests presented here, elevated HCN/HCO$^+$ ratios are not a reliable method of finding AGN, considering both the X-ray detections and upper limits.}
\end{itemize}

\acknowledgments
    The authors thank the anonymous referee for their constructive comments that have improved the clarity and quality of the paper.
    The authors also thank Masatoshi~Imanishi for comments on an earlier version of the manuscript.

GCP and DLJ acknowledge support from a NASA \nustar award 80NSSC17K0623.
GCP also acknowledges support from the University of Florida.
CR acknowledges support from the CONICYT+PAI Convocatoria Nacional subvencion a instalacion en la academia convocatoria a\~no 2017 PAI77170080.
T.D-S. acknowledges support from the CASSACA and CONICYT fund CAS-CONICYT Call 2018.
E.G-A is a Research Associate at the Harvard-Smithsonian Center for Astrophysics, and thanks the Spanish Ministerio de Econom\'{\i}a y Competitividad for support under project ESP2017-86582-C4-1-R.
KI acknowledges support by the Spanish MINECO under grant AYA2016-76012-C3-1-P and MDM-2014-0369 of ICCUB (Unidad de Excelencia 'Mar\'ia de Maeztu').
LB acknowledges support from National Science Foundation grant AST-1715413.
FEB acknowledges support from CONICYT-Chile (Basal AFB-170002) and the Ministry of Economy, Development, and Tourism's Millennium Science Initiative through grant IC120009, awarded to The Millennium Institute of Astrophysics, MAS.
ET acknowledges support from CONICYT-Chile grants Basal-CATA AFB-170002, FONDECYT Regular 1160999 and 1190818, and Anillo de ciencia y tecnologia ACT1720033.
KLL was supported by NASA through grants HST-GO-13690.002-A and HST-GO-15241.002-A from the Space Telescope Science Institute, which is operated by the Association of Universities for Research in Astronomy, Inc., under NASA contract NAS5-26555.
MPT acknowledges financial support from the State Agency for Research of the Spanish MCIU through the ``Center of Excellence Severo Ochoa'' award to the Instituto de Astrofísica de Andalucía (SEV-2017-0709) and through grant PGC2018-098915-B-C21.

A portion of this work was performed at the Aspen Center for Physics, which is supported by National Science Foundation grant PHY-1607611.
The authors thank the Sexten Center for Astrophysics (\url{http://www.sexten-cfa.eu}) where part of this work was performed.
This research has made use of the NASA/IPAC Extragalactic Database (NED) and NASA's Astrophysics Data System.

\facility{NuSTAR}

\software{ipython \citep{Perez2007},
numpy \citep{Vanderwalt2011},
scipy \citep{scipy-nmeth},
matplotlib \citep{Hunter2007},
Astropy \citep{astropy,Astropy2018},
Aplpy \citep{aplpy},
DESPOTIC \citep{Krumholz2014},
UCLCHEM \citep{Viti2004,Holdship2017},
CLOUDY \citep{Ferland2017}
}

\bibliography{ms}

\end{document}